\documentclass[preprint2]{proto}
\usepackage{times}
\newcommand{\refs}{\par\noindent\hangindent=1pc\hangafter=1}
\usepackage{fancyheadings}
\begin{document}

\lhead{}
\chead{}
\rhead{TO APPEAR IN PROTOSTARS AND PLANETS V}
\newcommand{\vdag}{(v)^\dagger}
\newcommand{\myemail}{chas@ipac.caltech.edu}
\newcommand{\etal}{{\it et al.\ }}

\title{Comparative Planetology and the Search for Life Beyond the Solar System}

\author{Charles A. Beichman}

\affil{Michelson Science Center, California Institute of Technology}
\email{chas@ipac.caltech.edu}

\author{Malcolm Fridlund}
\affil{ESTEC, European Space Agency}
\email{mfridlun@rssd.esa.int}

\author{Wesley A. Traub and Karl R. Stapelfeldt }
\affil{Jet Propulsion Laboratory, California Institute of Technology}
\email{wtraub@jpl.nasa.gov, krs@exoplanet.jpl.nasa.gov}

\author{Andreas Quirrenbach}
\affil{University of Leiden}
\email{quirrenb@strw.leidenuniv.nl}

\author{Sara Seager}
\affil{Carnegie Institute of Washington}
\email{seager@dtm.ciw.edu}

\begin{abstract}
\baselineskip = 11pt
\leftskip = 0.65in 
\rightskip = 0.65in
\parindent=1pc

{\small 

The study of planets beyond the solar system and the search for other habitable planets and life is just beginning. Ground-based (radial velocity and transits) and space-based surveys (transits and astrometry) will identify planets spanning a wide range of size and orbital location, from Earth-sized objects within 1 AU to giant planets beyond 5 AU, orbiting stars as near as a few parsec and as far as a kiloparsec. After this initial reconnaissance, the next generation of space observatories will directly detect photons from planets in the habitable zones of nearby stars. The synergistic combination of measurements of mass from astrometry and radial velocity, of radius and composition from transits, and the wealth of information from the direct detection of visible and mid-IR photons will create a rich field of comparative planetology. Information on proto-planetary and debris disks will complete our understanding of the evolution of habitable environments from the earliest stages of planet-formation through to the transport into the inner solar system of the volatiles necessary for life.

The suite of missions necessary to carry out the search for nearby, habitable planets and life requires a ``Great Observatories'' program for planet finding (SIM PlanetQuest, Terrestrial Planet Finder-Coronagraph, and Terrestrial Planet Finder-Interferometer/Darwin), analogous to the highly successful ``Great Observatories Program'' for astrophysics. With these new Great Observatories, plus the James Webb Space Telescope, we will extend planetology far beyond the solar system, and possibly even begin the new field of comparative evolutionary biology with the discovery of life itself in different astronomical settings. \\~\\~\\~}

\end{abstract}

\section{\textbf{STUDIES OF PLANETARY SYSTEMS AND THE SEARCH FOR LIFE}}

The search for habitable planets and life beyond Earth represents one of the oldest questions in natural philosophy, but one the youngest fields in astronomy. This new area of research derives its support among the scientific community and the general public from the fact that we are using 21$^{st}$ Century technology to address questions that were first raised by inquiring minds almost 2,500 years ago. Starting in 1995, radial velocity and, most recently, transit and microlensing studies have added more than 168 planets to the nine previously known in our own solar system. With steadily improving instrumentation, the mass limit continues to drop while the semi-major axis limit continues to grow: a 7.5 M$_\oplus$ planet orbiting the M star GL 876 at 0.02 AU (Rivera \etal 2005)  and a 4 M$_{Jup}$ planet orbiting 55 Cnc at 5.2 AU (Marcy \etal 2002) define boundaries which are sure to be eclipsed by newer discoveries. While we do not yet have the tools to find true solar system analogues or an Earth in the habitable zone of its parent star, evidence continues to accumulate (Marcy \etal 2005) that the number of potential Earths is large and that some could be detected nearby, if only we had the tools. In this article on space missions, we assess the prospects for the discovery and eventual characterization of planets of all sizes --- from gas-giants to habitable terrestrial analogues. Considerations of length necessarily make this discussion incomplete and we have omitted discussion of valuable techniques which do not by their nature, e.g. microlensing, lend themselves to follow-up observations relevant to physical characterization of planets and the search for life. 

\begin{table*}[tbp]
\tabletypesize{\scriptsize} 
\caption{Space Projects Presently Under Consideration \label{projtable}}
\begin{tabular}{lccc} 
&Approx. Sensitivity &Planet & Approximate \\
& Planet Size, Orbit, Dist. &Yield$^1$ & Launch Date\\
\hline
\multicolumn{4}{l}{\it Survey for Distant Planets (Transits)} \\
COROT& 	$2 R_\oplus$ at 0.05 AU (500 pc)&10s-100&2006\\
Kepler&	$1 R_\oplus$ at 1 AU (500 pc)&100s&2008\\
GAIA &  $10 R_\oplus$ at $<$0.1 AU &4,000&2012\\
JWST & 	$2 R_\oplus$ (100-500 pc) &250$^2$ &2013\\
\hline
\multicolumn{4}{l}{\it Determine Masses/Orbits (Astrometry)} \\
SIM& $ 3 M_\oplus$ at 1 AU (10 pc)&250&2012\\
GAIA& $ 30 M_\oplus$ at 1 AU ($<$200 pc)&10,000&2012\\
\hline
\multicolumn{4}{l}{\it Characterize Planets and Search for Life (Direct Detection)}\\
JWST&$ 1 R_{Jup}$ at $>30$ AU (50-150 pc) &250 young stars &2013\\
TPF-Coronagraph$^3$&$ 1 R_\oplus$ at 1 AU (15 pc)&250&2018\\
TPF-Interferometer/Darwin$^3$&$1 R_\oplus$ at 1 AU (15 pc)&250&2018\\
\end{tabular}
\tablenotetext{1}{Yield is highly approximate and assumes roughly 1 planet orbiting each star surveyed.}
\tablenotetext{2}{Approximate number of follow-up of ground-based, COROT and Kepler transit targets.}
\tablenotetext{3}{Parameters for TPF-C and TPF-I/Darwin are still being developed.}
\end{table*}
	
Papers in these proceedings describe many results from planet-finding experiments. With the exception of a few (very exciting) HST and Spitzer observations, these are drawn primarily from ground-based observations. This chapter on space activities necessarily focuses on future activities. While the space environment is highly stable and offers low backgrounds and an unobscured spectral range, the technology needed to take full advantage of the environment will take many years to develop and the missions to exploit the technology and the environment will be expensive to construct, launch and operate. But when the missions described here are completed, they will revolutionize our conception of our place in the Universe. 

\begin{figure*}
\epsscale{1.0}
\plotone{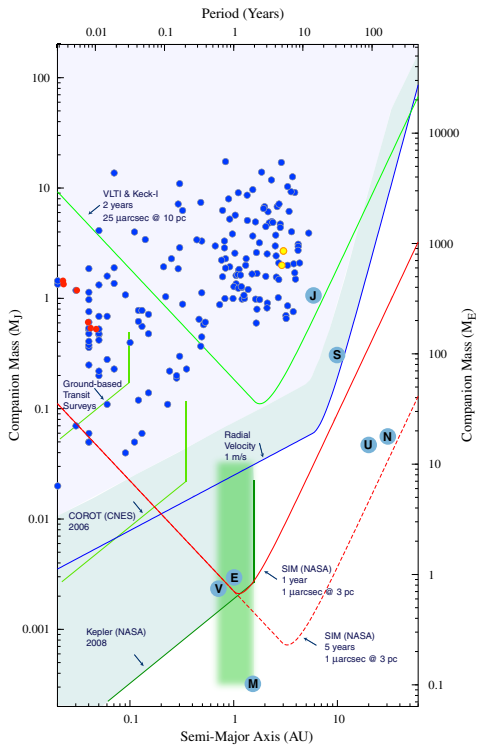}
\caption{\small A wide variety of space- and ground-based capabilities are summarized in this figure showing the detectability of planets of varying sizes and orbital locations. Space-based techniques become critical as one goes after terrestrial planets in the habitable zone. For a detailed description of this figure, see Lawson \etal  (2004).} \label{projfig}
\end{figure*}

Table~\ref{projtable} identifies the major space-based projects presently under consideration grouped by observing technique: transit photometry, astrometry, and direct detection. Figure~\ref{projfig} summarizes the discovery space of these projects as well as some ground-based activities in the Mass--Semi-Major Axis plane. COROT, Kepler and SIM will provide the many order-of-magnitude improvements in sensitivity and resolution relative to ground-based efforts needed to detect other Earths in the habitable zones of their parent stars using indirect techniques of transits, radial velocity and astrometry.

Once we have detected these planets via indirect means ($\S$2 and $\S$3), we argue below that we will need a number of missions to characterize these planets physically and to search for evidence of life in any atmospheres these planets may possess. Analogously to the improved knowledge gained about astrophysical phenomena from using all of NASA's four Great Observatories (the Hubble Space Telescope, the Compton Gamma Ray Observer, the Chandra X-ray Observatory, and the Spitzer infrared telescope) and ESA's Cornerstone missions (ISO and XMM), a comparable ``Great Observatories'' program for planet finding will yield an understanding of planets and the search for life that will greatly exceed the contributions of the individual missions. 

As discussed in $\S$2, the combination of transit photometry with COROT and Kepler, follow-up spectrophotometry from the James Webb Space Telescope (JWST) and follow-up radial velocity data gives unique information on planetary mass, radius, density, orbital location, and, in favorable cases, composition of the upper atmosphere. These data, available for large numbers of planets, will revolutionize our understanding of gas-giant and icy-planets. While less information will be available for smaller, rocky planets, a critical result of the transit surveys will be the frequency of Earth-sized planets in the habitable zone, $\eta_\oplus$ (Beichman 2000). Since the angular resolution and collecting area needed for the direct detection of nearby planets are directly related to the distance to the closest host stars, the value of $\eta_\oplus$ will determine the scale and cost of missions to find and characterize those planets.

Subsequent to the transit surveys, we will embark on the search for and the characterization of nearby planets, and the search for a variety of signposts of life. We will ultimately require three complementary datasets: masses via astrometry (SIM PlanetQuest, $\S$3); optical photons (TPF-Coronagraph, $\S$4); and mid-IR photons (TPF-Interferometer/Darwin, $\S$5). JWST will play an important role in follow-up activities looking at ground-based, COROT and Kepler transits; making coronagraphic searches for hot, young Jupiters; and studying proto-planetary and debris disks. The synergy between the planet finding missions, with an emphasis on studies of nearby, terrestrial planets and the search for life, is addressed in $\S$6. Studies of potential target stars are ongoing but will need to be intensified and their results collated to select the best targets ($\S$7). The ordering of these missions will be the result of the optimization of a highly non-linear function incorporating technical readiness, cost, and political and scientific support on two or more continents ($\S$8).

\section{TRANSITING PLANETS}

The age of comparative planetology is upon us with a growing
number of transiting giant planets---nine and counting. As they pass in front
of and behind their parent stars, the transiting planets' size---and
hence density, transmission spectrum, and thermal emission and albedo
can potentially be measured. This makes the group of transiting
planets the ones that can best be physically characterized---before
direct imaging is available. See Udry {\it et al.},  Charbonneau {\it et al.},  
and Marley \etal (all in these proceedings) for details on the
recent planet transit discoveries, observations, and interpretations.

The dedicated, space-based, transit survey missions --- Kepler and COROT --- will build upon the exciting ground-based transit detections of giant planets.
With very high-precision photometry enabled by the stable space
environment and lack of day/night and weather interruptions, COROT and
Kepler will push to planetary sizes as small as the Earth's. With
long-duration observing campaigns they will extend transit planet
discoveries to larger semi-major axes. JWST will similarly build on
the pioneering HST and Spitzer measurements of planetary atmospheres. The
legacy of Kepler and COROT combined with JWST will be to enable
comparative planetology on a wide range of planet types, 
encompassing a range of planet masses, temperatures, and host stars, 
before direct imaging of solar-system-aged planets is possible.

\bigskip
\bigskip
\noindent
\textbf{2.1 Prospects for Planet Transit Discoveries }
\bigskip

\noindent {\it HST and MOST.} The Hubble Space Telescope (HST) and the MOST (Microvariability \& OScillations of STars) microsatellite of the Canadian Space Agency have shown the promise of space-based transit studies. HST monitored 34,000 stars in the globular cluster 47 Tuc (Gilliand \etal 2000) continuously for 8.3 days. While 17 short-period giant planets were expected, none were found, suggesting that either low-metallicity or high stellar density interfere with planet formation or migration. More recently, HST ACS/WFC
monitored a field of 160,000 main sequence stars in the Galactic
bulge field for 7 days (Sahu \etal 2005). Over 100 transiting planets were
expected if the frequency of hot Jupiters in the Galactic bulge is
similar to that in the solar neighborhood. MOST, launched in June
2003, is a 15 cm telescope with a 350-700 nm broadband filter and a
part-per-million photometric accuracy capability for bright stars 
monitored for one month or more. MOST is not a transit survey
instrument, but has monitored four stars hosting known hot Jupiters
for 10 to 30 days. Of relevance to COROT and Kepler, MOST has put an
upper limit on the albedo of HD209458b of 0.15 (1$\sigma$) (Rowe \etal
2006) and is finding hints that host stars of the short-period
planets (hot Jupiters) may be too variable to detect the illumination 
phase curve (Walker \etal 2005).

\medskip
\noindent{\it COROT and Kepler.} COROT (COnvection, ROtation and planetary Transit) and Kepler are wide-field survey space telescopes designed to detect small transiting planets via  extremely high precision photometry. These telescopes will initiate the next generation of exoplanetary science by uncovering Neptune- (17 M$_\oplus$) to Earth-size planets around a range of stellar types. A large pool of this as yet unknown class of low-mass planets will provide the planet frequency and orbital distribution for insight into their formation and migration. The same group of planets will yield many objects
suitable for follow-up physical characterization.

COROT is a CNES/ESA mission to be launched in September 2006
(Baglin \etal 2003). COROT is a 27 cm telescope with a 3.5 deg$^2$
field of view. For the planet survey part of the program, five fields
containing approximately 12,000 dwarf stars in the range 11 $<V<$ 16.5 mag will be continuously monitored for 150 days. COROT will detect over 10,000
planets in the 1 to 5 $R_{\oplus}$ range within 0.3 AU assuming all stars have one such planet (Bord\'e \etal 2003). More realistic simulations show that COROT will detect about 100 transiting planets down to a size of 2 $R_{\oplus}$ around G0V stars and 1.1 $R_{\oplus}$ around M0V stars. See Gillon \etal (2005) and  Moutou \etal (2005) for details including the radius and semi-major axis distribution of expected transiting planets around different star types.

Kepler is a NASA Discovery mission (Borucki \etal 2003) to be launched in
June 2008. Kepler has a 0.95 m diameter mirror and an extremely wide
field of view---105 deg$^2$. Kepler will simultaneously monitor more
than 100,000 main sequence stars ($V<$14 mag) for its 4-year mission
duration. Kepler will find 50 transiting Earth-sized planets in the
0.5-1.5 AU range, if every star has 2 terrestrial planets (as the Sun
does). This number increases to 650 planets if most terrestrial
planets have a size of 2 $R_\oplus$. If Kepler finds few Earth-sized
planets it will come to the surprising and significant conclusion that
that Earth-size planets in Earth-like orbits are rare. Aside from
detecting Earth-sized planets in the habitable zone, Kepler will
advance the hot Neptune and hot Earth studies started by COROT, 
detecting up to hundreds of them down to a size as small as that of Mercury.

Both Kepler and COROT will produce exciting extrasolar giant planet science with tens of transiting giant planets with semi-major axes from 0.02-1 AU. Even giant planets in outer orbits---beyond 1 AU---can be detected with Kepler from single transit events at $8\sigma$. Follow-up radial velocity observations are required to confirm that the photometric dips are really due to planetary transits, as well as to measure the planetary masses and in some cases to determine the orbital period.

\bigskip
\noindent
\textbf{2.2 Physical Characterization of Transiting Planets }
\bigskip

The prospects for physical characterization of transiting planets
first with Spitzer and then with JWST are truly astonishing. By the
time JWST launches, one hundred or more transiting planets from
Jupiter down to Earth-sizes should be available for observation. While
the most favorable planets of all sizes will be at short semi-major
axes, a decent number of larger planets out to 1 AU will also be
suitable for density and atmosphere measurements. For transiting
giant planets, their albedo, moons and rings (Brown \etal 2001), and even
the oblateness and hence rotation rate (Seager \& Hui 2002; Barnes \& Fortney 2003) can potentially be measured. Some specific possibilities of physical
characterization are given below.

\medskip
\noindent {\it  Photometry.} JWST's NIRSpec is a high-resolution spectrograph from 0.7 to 5 $\mu$m. With its spectral dispersion and high cadence observing
NIRSpec will be capable of high-precision spectrophotometry on bright
stars. NIRSpec data can then be used in the same way that the HST STIS
spectral data for HD209458 was rebinned for photometry (Brown \etal 2001).
For example, at 0.7 $\mu$m\ JWST can obtain 35$\sigma$ transit detection for two interesting cases: an Earth-sized moon orbiting HD209458b (3 hour transit time at 47 pc) and a 1 AU Earth-sized planet orbiting a sun-like star at 300 pc\footnote{Gilliland in ``Astrobiology and JWST'' Report to NASA http://www.dtm.ciw.edu/seager/NAIAFG/JWST.pdf}. Kepler stars are about 300 pc distant---meaning JWST will be capable of confirming Kepler Earth-size planet candidates.

Planetary density is the key to the planetary bulk composition. With
precise radii from JWST and complementary radial velocity mass
measurements, densities of many planets can be determined, even for
super-Earth-mass planets close to the star. This will identify the
nature of many Neptune-mass and super-Earth-mass planets. Are they
ocean planets? Carbon planets? Small gaseous planets? Remnant cores of
evaporated giant planets? Or some of each? In this way JWST + Kepler/COROT 
will be able to provide insight into planet formation and migration of the
low-mass planets.

{\it Spectroscopy. } Planetary temperature is important for understanding planetary atmospheres and composition. Spitzer has initiated comparative
exoplanetology by measuring transiting hot Jupiters in the thermal
infrared using secondary eclipse (Deming \etal 2005, Charbonneau \etal 2005). Four transiting planets are being observed in 2005. Spitzer's
broad-band photometry from 3-8 $\mu$m, together with photometry at 14
and 24 $\mu$m will help constrain the temperatures and compositions
of hot Jupiters, including possibly their metallicity. The thermal
infrared phase variation of 7 hot Jupiters may also be detected this
year, providing clues about planetary atmospheric circulation in the
intense irradiation environment.

The JWST thermal IR detection capability can be explored by scaling
the Spitzer results. The 5-8 $\mu$m region is ideal for solar-type
stars because the planet-star contrast is high and the exo-zodiacal
background is low. For an estimate we can scale the TrES-1 5$\sigma$
detection at 4.5 $\mu$m, taking into account that JWST has 45 times the
collecting area of Spitzer, and assuming that the overall efficiency
is almost 2x times lower, giving an effective collecting area
improvement of $\sim$25 times. JWST will be therefore be able to
detect hot Jupiter thermal emission at an SNR of 25 around stars at
TrES-1's distance ($\sim$150 pc; a distance that includes most stars
from shallow ground-based transit surveys). Similarly, JWST can
detect a hot planet 5 times smaller than TrES-1, or down to 2 Earth
radii, for the same set of stars {\it assuming instrument systematics are
not a limiting factor}. Scaling with distance, JWST can detect hot
Jupiters around stars 5 times more distant than TrES-1 to SNR of 5, 
which includes all of the Kepler and COROT target stars. Beyond
photometery, JWST can obtain thermal emission spectra (albeit at a
lower SNR than for photometry for the same planet). Rebinning the
R=3,000 NIRSpec data to low-resolution spectra will enable detection of
H$_2$O, CO, CH$_4$, and CO$_2$.

Transiting planets too cold to be observed in thermal emission can
still be observed via transmission spectroscopy during primary planet
transit. Such planets include giant planets at all semi-major axes
from their stars. In visible and near-IR wavelengths, NIRSpec might
detect H$_2$O, CO, CH$_4$, Na, K, O$_2$ and CO$_2$. With NIRSpec
capabilities to a wavelength as short as 0.7 $\mu$m, JWST has the
potential to identify molecular oxygen at 0.76 $\mu$m (a sign of life
as we know it) in the outer atmosphere of a super-Earth-mass planet.

\section{ASTROMETRY}

\noindent
\textbf{3.1 Why Astrometry?}
\bigskip

The principle of planet detection with astrometry is similar to that behind the
Doppler technique: the presence of a planet is inferred from the motion of its
parent star around the common center of gravity. In the case of astrometry one
observes the two components of this motion in the plane of the sky; this gives
sufficient information to solve for the orbital elements without the $\sin i$
ambiguity plaguing Doppler measurements. In addition, the astrometric method
can be applied to all types of stars (independently of their spectral
characteristics), is less susceptible to noise induced by the stellar
atmosphere, and is more sensitive to planets with large orbital semi-major
axes. From simple geometry and Kepler's Laws it follows immediately that the
astrometric signal $\theta$ of a planet with mass $m_p$ orbiting a star with
mass $m_\ast$ at a distance $d$ in a circular orbit of radius $a$ is given by

$$
\theta = \frac{m_p}{m_\ast}\, \frac{a}{d} =\left(\frac{G}{4 \pi^2}
\right)^{1/3} \frac{m_p}{m_\ast^{2/3}} \, \frac{P^{2/3}}{d}$$

$$ = 3\, \mu{\rm as} \cdot \frac{m_p}{M_\oplus} \cdot \left(
\frac{m_\ast}{M_\odot}\right)^{-2/3} \left(\frac{P}{{\rm yr}}\right)^{2/3}
\left(\frac{d}{{\rm pc}}\right)^{-1} ~~~~(1). $$

This signature is represented in Fig.~\ref{projfig} which shows astrometric detection limits for a ground-based and space-based programs.

\bigskip
\noindent
\textbf{3.2 Astrometry from the Ground }
\bigskip

The best prospects for astrometric planet detection from the ground will be
offered by the development of narrow-angle dual-star interferometry (Shao \&
Colavita 1992; Quirrenbach \etal  1998; Traub \etal 1996), which is being pursued at the Palomar Testbed Interferometer (PTI; Muterspaugh \etal  2005), the Keck Interferometer (KI) and at ESO's Very Large Telescope Interferometer (VLTI; Quirrenbach \etal \ 2000). While the Earth's atmosphere imposes limits on ground-based interferometers, in favorable cases where a suitable reference star is available within $10''$, a precision of $10\, \mu$as can be reached. While adequate for gas giant planets, this limit excludes the use of ground-based facilities for searches for Earth analogues (cf.\ Fig.~\ref{projfig}). The scientific goals of ground-based projects thus include determining orbital inclinations and masses for planets already known from radial-velocity surveys, searches for giant planets around stars that are not amenable to high-precision radial-velocity observations, and a search for large rocky planets around nearby low-mass stars.

\bigskip
\noindent
\textbf{3.3 SIM PlanetQuest: Nearby Terrestrial Planets }
\bigskip

NASA's Space Interferometry Mission (SIM PlanetQuest) will push the precision
of astrometric measurements far beyond the capabilities of any other project
currently in existence or under development. SIM, to be launched in 2012, will
exploit the advantages of space to perform a diverse astrometric observing
program, e.g., Unwin (2005) and Quirrenbach (2002). SIM consists of a single-baseline interferometer with 30\, cm telescopes on a 9\, m baseline. SIM
is a pointed mission so that targets can be observed whenever there is a
scientific need, subject only to scheduling and solar exclusion angle
constraints. Additionally, the integration time can be matched to the desired
signal-to-noise ratio enabling observation of very faint systems.

In its ``narrow-angle'' mode (i.e., over a field of about $1^\circ$), SIM will
provide an accuracy of $\sim 1\, \mu$as for each measurement. SIM PlanetQuest
will carry out a high-precision survey of $250$ nearby stars reaching down
to 1 to 3 M$_\oplus$ (depending on stellar mass and distance) and a
less sensitive survey of some 2,000 stars establishing better statistics on
massive planets in the Solar neighborhood. In addition, SIM will observe a
sample of pre-main-sequence stars to investigate the epoch of planet formation.

It is very likely that SIM PlanetQuest will discover the first planets in the
habitable zone around nearby stars. True Earth analogues are just within reach.
With 200 visits at a single-measurement precision of 1\, $\mu$as, astrometric
signatures just below 1\, $\mu$as constitute secure planet detections, with a
false-alarm probability of only 1\%. This means that planets with 1\, M$_\oplus$
in 1\, AU orbits can be discovered around seven nearby G and K dwarfs; planets
twice as massive would be found around 28 G and K stars. Assuming $\eta_\oplus
= 0.1$, SIM would have a $\sim 50\%$ chance of finding at least one
1\, M$_\oplus$ / 1\, AU planet, and a $\sim 95\%$ chance of discovering at least one 2\, M$_\oplus$ planet in a 1\, AU orbit. 

One should note that the astrometric signature of the Earth, 450~km or 1/1,500 R$_\odot$, is several times larger than the motion of the photocenter of the Sun induced by spots. Starspots are not expected to contribute significantly to the noise for planet searches around Sun-like stars. Although starspots are a cause for  concern for more active types of star, their effects are less than the radial velocity noise associated with young stars. SIM astrometry will be able to  find gas giant planets within a few AU of T Tauri stars for the first time.

\bigskip
\noindent
\textbf{3.4 GAIA: A Census of Giant Planets }
\bigskip

The European Space Agency is planning to launch an astrometric satellite, 
GAIA, in roughly the same time frame as SIM. GAIA's architecture builds on the
successful Hipparcos mission (Lindegren \& Perryman 1996; Perryman \etal \
2001). Unlike SIM, GAIA will be a continuously scanning survey instrument with
a large field of view, which will cover the whole sky quite uniformly, observing each star hundreds of times over its 5 year mission. Among the many scientific results of the GAIA mission will thus be a complete census of stellar and sub-stellar companions down to the accuracy limit of the mission. This limit depends on the magnitude and color of each star: for a G2V star the expected accuracy, expressed as the parallax error at the end of the mission is 7\, $\mu$as at $V=10$ mag and 25\, $\mu$as at $V=15$ mag. Very roughly, the corresponding {\it single measurement accuracy} relevant to planet detection will be 70\, $\mu$as at $V=10$ mag and 250\, $\mu$as at $V=15$ mag, assuming 100 measurements per star over the course of the mission. GAIA is thus expected to detect some 10,000 Jupiter-like planets in orbits with periods ranging from 0.2 to 10 years, out to a typical distance of 200\, pc. Around nearby stars, the detection limit will be about 30\, M$_\oplus$. In addition, over 4,000 transiting ``hot Jupiters'' will be detected in the GAIA photometry. GAIA will thus provide complementary information to SIM, on the incidence of planetary systems as a function of metallicity, mass and other stellar properties.

\section{CORONAGRAPHY}

\noindent
\textbf{4.1 Ground-based Coronagraphy}
\bigskip

The search for faint planets located close to a bright host star requires specialized instrumentation capable of achieving contrast ratios of 10$^{-9}-10^{-10}$ at subarcsecond separations. The most common approach is a Lyot coronagraph, where direct starlight is occulted at a first focal plane; diffraction from sharp edges in the entrance pupil is then suppressed with a Lyot mask in an intermediate pupil plane; and finally, imaging at a second focal plane occurs with greatly reduced diffraction artifacts. In addition to diffraction control, good wavefront quality (Strehl ratio) and wavefront stability are needed to achieve high image contrast. 

Future extremely large ground-based telescopes will complement space 
coronagraphy through studies of young or massive Jovian planets. Contrast 
levels of 10$^{-7}$ or perhaps 10$^{-8}$ may be achievable at subarcsec separations through future developments in extreme adaptive optics. However, achieving the higher contrasts needed to detect Earths would require working at levels of $10^{-3}$ and $10^{-4}$ below the level of the residual background, many orders of magnitude greater than has been demonstrated to date (Dekany \etal  2006; Chelli 2005; Beuzit \etal , this volume). As discussed below, the TPF-Coronagraph will greatly reduce the residual stellar background by taking advantage of a stable space platform and the absence of a variable atmosphere. Furthermore, the study of weak atmospheric features (including key biomarkers) on distant planets will be straightforward from space, but could be compromised by observing at relatively low spectral resolution through the Earth's atmosphere.

\bigskip
\noindent
\textbf{4.2 High Contrast Imaging from Space}
\bigskip

Ground-based adaptive optics systems and three Hubble Space Telescope (HST) instruments using simple coronagraphs have achieved valuable results on circumstellar disks and substellar companions at moderate image contrast 
(see Beuzit \etal  and M\'enard \etal , this volume; Schneider \etal  2002; 
Krist \etal  2004, and references therein). However, only by taking full advantage of stable, space-based platforms can the necessary starlight rejection ratios be achieved. To date, no space astronomy mission has been designed around the central goal of very high contrast imaging to enable detection of extrasolar planets. The HST coronagraphs were low priority add-ons to general-purpose
instruments, and achieved only partial diffraction control and image
contrasts of ~10$^{-4}$-10$^{-5}$. 

JWST, expected to launch in 2013, will offer coronagraph modes for its two infrared imaging instruments. Even though JWST's segmented primary mirror is poorly suited to high contrast coronagraphy, JWST's coronagraphs can access the bright 4.8 $\mu$m emission feature expected in the spectra of giant planets and brown dwarfs enabling the detection of substellar companions at contrast ratios of 10$^{-5}$-10$^{-6}$. Detailed performance estimates show that JWST should be able to detect warm planets in nearby young stellar associations at radii $>$ 0.7$^{\prime\prime}$, and perhaps even a few old (5 Gyr) Jupiter-mass planets around nearby, late-M type stars (Green \etal 2005b). 

The key to achieving such high contrasts is precision wavefront sensing and control. After a coronagraph suppresses the diffracted light in a telescope, the detection limits are governed by how much light is scattered off surface imperfections on the telescope primary mirror. In HST, these imperfections produce a background field of image speckles that is more than $1000 \times$ greater than the expected brightness of a planet in visible light. At the time HST was designed, the only way to reduce these imperfections would have been to polish the primary mirror to 30$\times$ better surface accuracy - infeasible then and now. But during the 1990s, another solution became available: use deformable mirrors (DMs) to actively correct the wavefront. The needed wavefront
quality can be achieved by canceling out primary mirror surface figure 
errors with a DM - exactly as Hubble's spherical aberration was corrected, 
but now at higher precision and higher spatial frequencies.

Over the past 5 years, the community has been developing deformable mirror  technology, innovative pupil masks, and instrument concepts to enable very high contrast imaging (Green et~al. 2005a; Kuchner and Traub 2002; Vanderbei et~al. 2004; Guyon et~al. 2005; Traub and Vanderbei 2005; Shao et~al. 2004). At the Jet Propulsion Laboratory, the High Contrast Imaging Testbed (HCIT) has been developed and consists of a vacuum-operated optical bench with a Lyot coronagraph and a 64$\times$64 format deformable mirror. New wavefront control algorithms have been developed. As of mid-2005, HCIT had {\it demonstrated} 10$^{-9}$ contrast in 785 nm narrowband light, at a distance of 4$\lambda$/D from a simulated star image (Green \etal  2005a) --- a huge improvement over the previous state of the art. While additional progress is needed to achieve the requirement of 10$^{-10}$ contrast in broadband light, these results are extremely encouraging evidence that a coronagraphic version of the Terrestrial Planet Finder mission (TPF-C) is feasible. NASA is now defining a TPF-C science program and mission design concept.

\bigskip
\noindent
\textbf{4.3 TPF-C Configuration }
\bigskip

\begin{figure*}
\epsscale{1.2}
\plotone{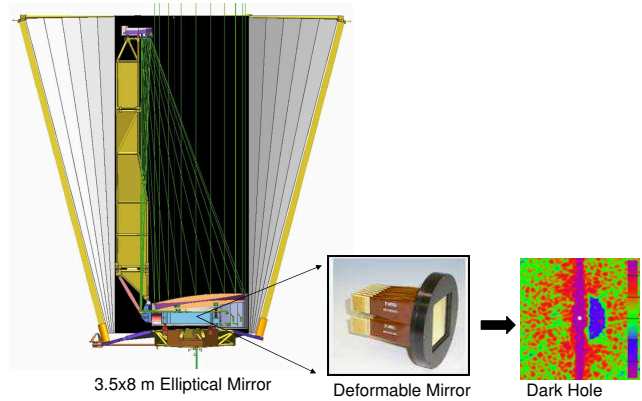}
\caption{\small Left) An artist's concept for TPF showing the sunshield surrounding the 3.5x8 m primary mirror; middle) a picture of the deformable mirror which is the key development for wavefront control; right) a ``dark hole'' is created when the deformable mirror is adjusted to take out wavefront errors in the optical system. In this rectangular area the contrast demonstrated in the laboratory has reached $10^{-9}$ starting at a field angle of $4\lambda$/D, within a factor of 10 of that needed to detect planets.} \label{tpfc}
\end{figure*}

The configuration of the TPF-C observatory is driven by the specialized 
requirements of its high contrast imaging mission (Figure~\ref{tpfc}). To enable searches of the habitable zone in at least 100 nearby FGK stars an 8 meter aperture is  needed. A monolithic primary mirror is essential to achieving the needed wavefront quality and maximizing throughput, but a conventional circular
8 m primary cannot be accommodated in existing rocket launcher shrouds.
The unconventional solution is an 3.5 m $\times$ 8 m elliptical primary, 
tilted on-end within the rocket shroud for launch, with a deployed secondary
mirror. The telescope has an off-axis (unobscured) design, to minimize 
diffraction effects and maximize coronagraph throughput. To provide very 
high thermal stability, the telescope is enclosed in a multi-layer V-groove 
sunshade and will be operated in L2 orbit. The large sunshade dictates
a solar sail to counterbalance radiation pressure torques on the spacecraft.
Milliarcsec-level pointing will be required to maintain alignment of the
bright target star on the coronagraph occulting spot.

TPF-C's main instrument will be a coronagraphic camera/spectrometer operating 
over the wavelength range 0.4 $< \lambda <$ 1.1 $\mu$m. It will have
a very modest field of view, perhaps 20$^{\prime\prime}$ in diameter, 
with a corrected high contrast dark field extending to a radius of
$\sim1^{\prime\prime}$ from the bright central star. The science camera
will also serve as a wavefront sensor to derive the adaptive correction
settings for the DM which must be set and maintained to sub-angstrom accuracy to achieve the required contrast. The camera will provide multispectral imaging, at resolutions from 5-70, in order to separate planets from residual stellar speckles, and to characterize their atmospheres. Spectral features of particular interest for habitability are the O$_2$ A band at 0.76 $\mu$m, an H$_2$O band at 0.81 $\mu$m, and the chlorophyll red edge beyond 0.7 $\mu$m.

Terrestrial exoplanets are very faint targets, with typical V$>$ 29 mag.
Even with a collecting area five times greater than HST's, exposure times of 
order a day will be needed to detect them. Repeat observations at multiple 
epochs will be needed to mitigate against unfavorable planetary elongations 
or illumination phases. Each of the target stars must therefore be searched 
for roughly a week of cumulative integration time. In systems where candidate 
planets are detected, follow-up observations to establish common proper 
motion, measure the planet's orbital elements, and spectroscopic 
characterization would add weeks of additional integration time per target. 
Accompanying giant planets and zodiacal dust should be readily detected 
in the systems selected for terrestrial planet searches. 

\section{INTERFEROMETRY}

While COROT and Kepler will tell us about the general prevalence of terrestrial planets, we need to study planets in their habitable zones around nearby (25 pc) solar type stars in a large enough sample (150 to over 500) to make statistically significant statements about habitable planets and the incidence of life. In order to cover a range of spectral types, metallicities, and other stellar properties, we would like to include many F, G, K and some M dwarfs. As described in $\S$4, the angular resolution of a space-based coronagraph is limited to about 60 mas ($4\lambda/D$ at 0.7 $\mu$m) by the size of largest monolithic telescope one can launch (roughly an $\sim8$ m major axis). Beyond about 15 pc such a coronagraph will be limited to searches of the habitable zones around more luminous F-type stars. An interferometer with each telescope on a separate spacecraft suffers no serious constraints on angular resolution and would be able to study stars over a large span of distances and luminosities with a consequently larger number and greater variety of potential targets. 

Nulling (or destructive) interferometry uses an adaptation of the classical Michelson interferometry currently being developed at ground based sites like the Keck Interferometer  and ESO's Very Large Telescope Interferometer. Interferometers operating in the mid-infrared offer a number of advantages over other systems:

\begin{itemize}

\item Inherent flexibility which allows the observer to optimize the angular resolution to suit a particular target star.

\item A star-planet contrast ratio that is only $10^{-6} - 10^{-7}$ at 10 $\mu$m, roughly a factor of 1,000 more favorable than at visible wavelengths.

\item A number of temporal and spatial chopping techniques to filter out optical and mechanical instabilities, thereby relaxing some difficult requirements on the optical system. 

\item The presence of deep, broad spectral lines of key atmospheric tracers that can be observed with low spectral resolution.

\end{itemize}

There are, of course, disadvantages to a mid-IR system, including the need for cryogenic telescopes to take advantage of low space background; the complexity of multi-spacecraft formation flying; and the complexity of signal extraction from interferometric data compared to more direct coronagraphic imaging.

ESA and NASA are investigating interferometric planet finding missions and investing heavily in key technologies to understand the tradeoffs between different versions of the interferometer and, more generally, with coronagraphs.

\bigskip
\noindent
\textbf{5.1 Nulling Interferometry }
\bigskip

In a nulling interferometer the outputs of the individual telescopes are combined after injecting suitable phase differences (most simply a half-wavelength) so that the on-axis light is extinguished while, at the same time, slightly off-axis light will be transmitted. By rotating the interferometer or by using more than two telescopes in the system one can sweep around the optical axis with high transmission while constantly obscuring the central object. In this manner, first proposed by Bracewell (1978), one can achieve the very high contrast ratios needed to detect a planet in the presence of its parent star. The depth and the shape of the null in the center depend critically on the number of telescopes and the actual configuration (Angel \& Woolf 1997). While better angular resolution is desirable to probe closer to the star, at high enough resolution the central stellar disk becomes resolved and light leaks out of the central null. This leakage is one of the noise sources in a nulling interferometer designed to search for terrestrial planets. When full account is taken of the leakage and other noise sources including local zodiacal emission, telescope background, and zodiacal light from dust orbiting the target star, the performance of a nulling interferometer using 3-4 m telescopes is well matched to the study of terrestrial planets around hundreds of stars (Beichman \etal 1998; Mennesson \etal 2004).

Excellent technical progress has been made in both the US and Europe on the key ``physics'' experiment of producing a broad-band null. Depths less than 10$^{-6}$ have been achieved in a realistic 4 beam configuration in the laboratory (Martin \etal  2003). Operational nulling systems are being deployed on both the Keck and Large Binocular Telescope Interferometers (Serabyn \etal  2004; Herbst et al 2004).

\bigskip
\noindent
\textbf{5.2 Darwin and the Terrestrial Planet Finder }
\bigskip

ESA's Darwin and NASA's TPF-I are currently foreseen to be implemented on 4-5 spacecraft flying in a precision formation. The system would consist of 3 or 4 telescopes each of 3-4 m diameter and flying on its own spacecraft. An additional spacecraft would serve as a beam combiner. When searching nearby stars out to 25 pc, the distance between the outermost telescopes would be roughly 100 m. The operating wavelength would be 6 to 17 $\mu$m (possibly as long as 20 $\mu$m) where the contrast between parental star and Earth-like planet is most favorable and where there are important spectral signatures characterizing the planets. ESA and NASA are collaborating on TPF-I/Darwin under a Letter of Agreement that calls for joint science team members, conferences and workshops as well as for periodic discussions on technology and the various configurations under study individually by the two space agencies. The common goal is to implement one interferometric mission, since complexity and cost indicate the necessity for a collaborative approach. Systems utilizing 1-2 m telescopes could also be considered if it were known in advance, e.g. from COROT and Kepler, that Earths in the habitable zone were common so that one could limit the search to 10 pc instead of 25 pc to detect a suitable number of planets.

\section{SYNERGY AMONG TECHNIQUES FOR NEARBY PLANETS}

The major missions discussed in this chapter approach the search for terrestrial planets from different perspectives: COROT and Kepler for transits ($\S$2); SIM using astrometry ($\S$3); TPF-C directly detecting visible photons  ($\S$4); and TPF-I/Darwin directly detecting mid-infrared photons ($\S$5). In this section we argue that all these perspectives are needed to determine habitability and search for signs of life. The synergy between the transit missions and JWST has already been discussed ($\S$2). In this section we focus on investigations of nearby stars and the potential for SIM, TPF-C, and TPF-I/Darwin to determine important physical parameters either individually or in combination (Table~\ref{synergy}). {\it The cooperative aspects are discussed below in italics.} We anticipate that the strongest, most robust statements about the characteristics of extrasolar planet will come from these cooperative measurements.

\begin{table*}[tbp]
\tabletypesize{\scriptsize} 
\caption{Measurement Synergy for TPF-C, TPF-I/Darwin and SIM \label{synergy}}
\begin{tabular}{lccc} \\ 
&SIM&TPF-C&TPF-I\\
\hline
{\it Orbital Parameters} & & & \\
\ \ \	Stable orbit in habitable zone&{\bf Measurement}&{\bf Measurement}&{\bf Measurement}\\
{\it Characteristics for Habitability}& & & \\
\ \ \	Planet temperature&Estimate&Estimate&{\bf Measurement}\\
\ \ \	Temperature variability due to eccentricity &{\bf Measurement}&{\bf Measurement}&{\bf Measurement}\\
\ \ \	Planet radius&{\it Cooperative}&{\it Cooperative}&{\bf Measurement}\\
\ \ \	Planet albedo&{\it Cooperative}&{\it Cooperative}&{\it Cooperative}\\
\ \ \	Planet mass&{\bf Measurement}&Estimate&Estimate\\
\ \ \	Surface gravity&{\it Cooperative}&{\it Cooperative}&{\it Cooperative}\\
\ \ \	Atmospheric and surface composition&{\it Cooperative}&{\bf Measurement}&{\bf Measurement}\\
\ \ \	Time-variability of composition&&{\bf Measurement}&{\bf Measurement}\\
\ \ \	Presence of water&&{\bf Measurement}&{\bf Measurement}\\
{\it Solar System Characteristics}& & & \\
\ \ \	Influence of other planets, orbit coplanarity&{\bf Measurement} & Estimate& Estimate\\
\ \ \	Comets, asteroids, and zodiacal dust&&{\bf Measurement}&{\bf Measurement}\\
{\it Indicators of Life}& & & \\
\ \ \	Atmospheric biomarkers&&{\bf Measurement}&{\bf Measurement}\\
\ \ \	Surface biosignatures (red edge of vegetation)&&{\bf Measurement}&\\
\end{tabular}
\tablenotetext{1}{{\bf ``Measurement''} indicates a directly measured quantity from a mission; ``Estimate'' indicates that a quantity that can be estimated from a single mission; and {\it ``Cooperative''} indicates a quantity that is best determined cooperatively using data from several missions.}
\end{table*}

\bigskip
\noindent
\textbf{6.1 Stable Orbit In Habitable Zone}
\bigskip

Each mission can measure an orbit and determine if it lies within the habitable zone (where the temperature permits liquid water on the surface of the planet). SIM does this by observing the wobble of the star and calculating where the planet must be to cause that wobble. TPF-C and TPF-I/Darwin do this by directly imaging the planet and noting how far it appears to be from the star. The missions work together and separately to determine orbital information:

\begin{itemize}

\item SIM detections of planets of a few Earth masses would provide TPF-C and TPF-I/Darwin with targets to be characterized and the optimum times for observing them, thus increasing the early-mission characterization yield of TPF-C or TPF-I/DARWIN.

\item Where SIM finds a planet, of any mass, in almost any orbit, TPF-C and TPF-I/Darwin will want to search as well, because we expect that planetary multiplicity may well be the rule (as in our Solar System). Thus SIM will help TPF-C and TPF-I/Darwin to prioritize likely target stars early in their missions.

\item For stars where SIM data suggest that planets exist below SIM's formal detection threshold, TPF-C or TPF-I/Darwin could concentrate on those stars to either verify or reject the detection. Such verification would lower the effective mass detection threshold for planets with SIM data. 

\item All three missions can detect several planets around a star, within their ranges of sensitivity. Thus there may be a planet close to the star that SIM can detect, but is hidden from TPF-C. Likewise there may be a distant planet that TPF-C or TPF-I/Darwin can detect, but has a period that is too long for SIM. For the more subtle issue of whether the planets have orbits in or out of the same plane, SIM will do the best job. {\it In general, each of the three missions will detect some but not necessarily all of the planets that might be present in a system, so the combination will deliver a complete picture of what planets are present, their masses, their orbits, and how they are likely to influence each other over the age of the system, including co-planarity. }

\end{itemize}

\bigskip
\noindent
\textbf{6.2 Gross Physical Properties of Planets }
\bigskip

\noindent {\it Planet temperature.} A planet's effective temperature can be roughly estimated by noting its distance from its star and assuming a value for the albedo. TPF-C can estimate the temperature by noting the distance and using planet color to infer its albedo by analogy with planets in our Solar System. TPF-I/Darwin can observe directly the thermal infrared emission continuum at several wavelengths (i.e., infrared color) and use Planck's law to calculate the effective temperature. For a planet like Venus with a thick or cloudy atmosphere, the surface temperature is different from the effective temperature, but might still be inferred from a model of the atmosphere. {\it With all three missions combined, the orbit, albedo, and greenhouse effect can be estimated, and the surface temperature as well as temperature fall-off with altitude can be determined cooperatively and more accurately than with any one mission alone.}

\medskip\noindent {\it Temperature Variability due to Distance Changes.} Each mission alone can observe the degree to which the orbit is circular or elliptical, and thereby determine if the temperature is constant or varying. In principle TPF-C and TPF-I/Darwin can tell whether there is a variation in color or spectrum at different points in the planet's orbit due, perhaps, to a tilt of the planet's axis which would lead to a seasonal temperature variability. {\it The measurement of a terrestrial planet's orbital eccentricity using combined missions (SIM plus TPF-C and/or TPF-I) can be much more accurate than from any one mission alone, because complementary sensitivity ranges in planet mass and distance from star combine favorably. SIM gives eccentricity data that aids TPF-C and TPF-I/Darwin in selecting optimum observation times for measuring planet temperature, clouds, and atmospheric composition. }

\medskip\noindent {\it Planet Radius.} SIM measures planet mass from which we can estimate radius to within a factor of 2 if we  assume a value for the density (which in the Solar System spans a factor of 8). TPF-C measures visible brightness, which along with an estimate of albedo, can give a similarly rough estimate of radius. A TPF-C color-based estimate of planet type can give a better estimate of radius. TPF-I/Darwin measures infrared brightness and color temperature which, using Planck's law, gives a more accurate planet radius. {\it Planet radius and mass, or equivalently density, is very important for determining the type of planet (rocks, gas, ice, or combination), its habitability (solid surface or not; plate tectonics likely or not), and its history (formed inside or outside of the ice-line). With SIM's mass, and one or both TPF brightness measurements, we can dramatically improve the estimate of planet radius.}

\medskip\noindent {\it Planet Albedo.} The albedo controls the planet's effective temperature which is closely related to its habitability. SIM and TPF-C combined can estimate possible pairs of values of radius and albedo, but cannot pick which pair is best (see above). We can make a reasonable estimate of albedo by using TPF-C to measure the planet's color, then appealing to the planets in our Solar System to convert a color to an absolute albedo. By adding TPF-I/Darwin measurements we can determine radius (above), then with brightness from TPF-C we can compute an accurate albedo. {\it SIM and TPF-C together give a first estimate of planet albedo. Adding TPF-I/Darwin gives a conclusive value of albedo, and therefore effective temperature and potential habitability.}

\medskip\noindent {\it Planet Mass.} SIM measures planet mass directly and accurately. TPF-C and TPF-I/Darwin depend entirely on SIM for a true measurement of planet mass. If TPF-C and TPF-I/Darwin do not have a SIM value for planet mass, then they will use theory and examples from our Solar System to estimate masses (see above). {\it SIM plus TPF-C and TPF-I/Darwin are needed to distinguish among rock-, ice-, and gas-dominated planet models, and to determine with confidence whether the planet could be habitable.}

\medskip\noindent {\it Surface Gravity.} The planet's surface gravity is calculated directly using mass from SIM and radius from TPF-C and TPF-I/Darwin (see above). {\it Surface gravity governs whether a planet can retain an atmosphere or have plate tectonics (a crucial factor in Earth's evolution). Cooperative measurements are the only way to obtain these data.}

\medskip\noindent {\it Atmosphere and Surface Composition.} The TPF missions are designed to measure a planet's color and spectra from which we can determine the composition of the atmosphere and surface. For the atmosphere, TPF-C can measure water, molecular oxygen, ozone, the presence of clouds for a planet like the present Earth, and in addition it can measure carbon dioxide and methane for a planet like the early Earth or a giant planet. For the surface TPF-C can measure vegetation using the "red edge" effect (see below). TPF-I/Darwin will add to this suite of observations by measuring carbon dioxide, ozone, water, methane, and nitrous oxide using different spectroscopic features, and in general probing a different altitude range in the atmosphere. SIM is important to this interpretation because it provides planet mass, crucial to interpreting atmospheric measurements.

Both TPF-C and TPF-I/Darwin are needed in order to determine whether a planet is habitable, because they make complementary observations, as follows (assuming an Earth-like planet). Ozone has a very strong infrared (TPF-I) feature, and a weak visible (TPF-C) one, so if ozone is abundant, both can be used to extract the abundance. If ozone is only weakly present, then only the TPF-I/Darwin feature will be useable. Water as seen by TPF-C will be in the lower atmosphere of the  planet, but as seen by TPF-I/Darwin it will be in the upper atmosphere; together both give a more complete picture of the atmosphere. Methane and carbon dioxide could be detected by TPF-I/Darwin at levels similar to present day concentrations in the Earth's atmosphere. Methane and carbon dioxide, in large amounts (as for the early Earth), can be detected by TPF-C. For large amounts of methane or carbon dioxide, TPF-I/Darwin will see mainly the amount in the upper atmosphere, but TPF-C will see mainly the amount in the lower atmosphere, so both are needed for a complete picture. In addition to these "overlap" topics, only TPF-C can potentially measure oxygen, vegetation, and the total column of air (Rayleigh scattering); likewise only TPF-I/Darwin can measure the effective temperature. TPF-I's wavelength coverage includes a spectral line of nitrous oxide, a molecule strongly indicative of the presence of life. Unfortunately, there is only a small hope of detecting this biomarker at low spectral resolution. {\it In short, SIM is needed for planet mass, TFP-C and TPF-I/Darwin are needed to characterize the atmosphere for habitability, and all three are needed to fully characterize the planet.}

\medskip\noindent {\it Temporal Variability of Composition.} Both TPF-C and TPF-I/Darwin potentially can measure changes in color and the strengths of spectral features as the planet rotates. These changes can tell us the length of day on the planet, and can indicate the presence of large oceans or land masses (with different reflectivities or emissivities, by analogy to Earth). Superposed on this time series of data could be random changes from weather patterns, possibly allowing the degree of variability of weather to be measured. {\it The TPF missions can potentially measure variability of composition over time, which we know from our Earth to be an indicator of habitability.}

\medskip\noindent {\it Presence of Water.} Both TPF-C and TPF-I/Darwin have water absorption features in their spectra, so if water vapor is present in the atmosphere, we will be able to measure it. However habitability requires liquid water on the surface, which in turn requires a solid surface as well as a temperature that permits the liquid state; only with the help of a value of mass from SIM will we be able to know the radius, and when TPF-I/Darwin is launched, the temperature. {\it To know whether liquid water is present on the surface of a planet, we need mass data from SIM, and spectroscopic data from TPF-C or TPF-I/DARWIN.}

\bigskip
\noindent
\textbf{6.3 Biomarkers }
\bigskip

The simultaneous presence of an oxidized species (like oxygen or ozone) and a reduced species (like methane) is considered to be a sign of non-equilibrium that can indicate indirectly the presence of life on a planet. The presence of a large amount of molecular oxygen, as on the present Earth, may also be an indirect sign of life. In addition since water is a prerequisite for life, as we consider it here, the presence of liquid water (indicated by water vapor and an appropriate temperature) is needed. Together these spectroscopically-detectable species are our best current set of indicators of life on a planet. {\it These markers will be measured exclusively by TPF-C and TPF-I, but to know that we are observing an Earth-like planet will require SIM data on mass. If we do (or do not) find biomarkers, we will certainly want to know how this is correlated with planet mass.}

The ``red edge'' of vegetation is a property of land plants and trees whereby they are very good reflectors of red light just beyond the long-wavelength limit of our eyes. This is a useful feature for measuring plant cover on Earth. If extrasolar planets have developed plant life like that on Earth, and if the planet is bright, has few clouds, and a lot of vegetated land area, then we may use this feature to detect living vegetation. {\it As for other biomarkers (above), we will want to correlate the presence of vegetation with the planet mass, requiring SIM as well as TPF-C.}

\begin{table}[h]
\tabletypesize{\scriptsize} 
\caption{ Key Target Star Properties \label{starproptable}}
\begin{tabular}{ll} 
\hline
Stellar Age & Evolutionary Phase \\
Spectral Type & Mass \\
Variability & Metallicity\\
Distance & Galactic Kinematics \\
Multiplicity & Giant Planet Companions\\
Exozodiacal Emission & Background Confusion\\
Position in Ecliptic & Position in Galaxy\\
\hline
\end{tabular}
\end{table}

\section{TARGET STARS}
\noindent
\textbf{7.1 Stars Suitable for SIM, TPF-C and TPF-I/Darwin }
\bigskip

Table~\ref{starproptable} suggests just a few of the considerations that will go into choosing the best targets for SIM, TPF-C and TPF-I/Darwin. Some of these are of a scientific nature, e.g. age, spectral type, and metallicity (which seems to be highly correlated with the presence of gas giant planets), while other constraints are more of an engineering nature: zodiacal emission, ecliptic latitude, binarity, variability, or the presence of confusing background objects. An extensive program of observation and data gathering must be an essential part of preparation for the planet finding program. A report entitled {\it Terrestrial Planet Finder Precursor Science} (Lawson et~al. 2004) describes a complete roadmap for the acquisition and assessment of relevant data. NASA has begun development of the Stellar Archive and Retrieval System (StARs) to provide a long term repository and network accessible database for this effort. A parallel and complementary effort is currently being developed in Europe.

\begin{figure*}
\epsscale{1.8}
\plotone{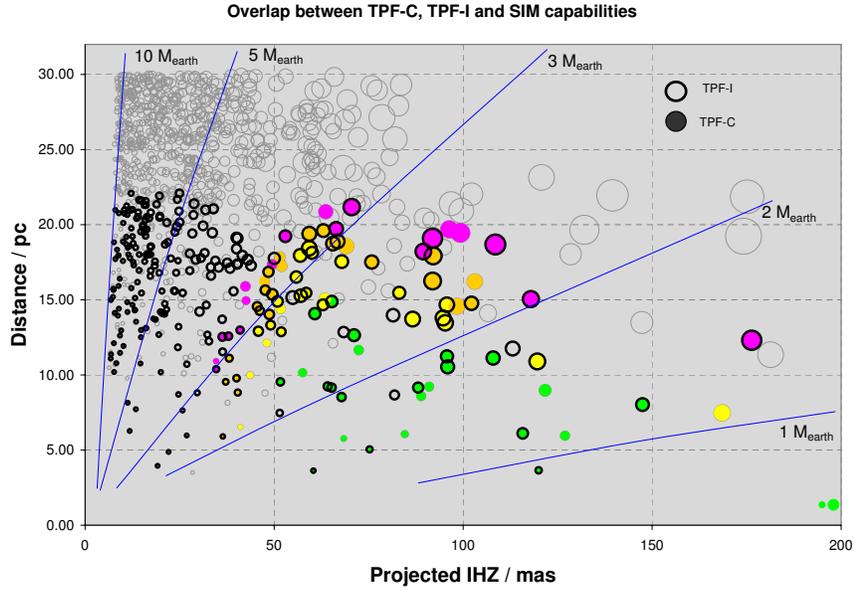}
\vskip -2.0in
\caption{\small Potential targets for SIM, TPF-C, and TPF-I/Darwin are shown in terms of stellar distance versus the angular extent of the inner edge of the habitable zone (IHZ, which is set to be the orbit of Venus scaled by the square root of the stellar luminosity). Stellar diameter is indicated by size of the symbol. High priority  TPF-C targets are shown with filled symbols shaded to denote different probabilities of completeness (Brown 2005) in a survey for Earth-sized planets in the center of the habitable zone (green denotes highest completeness ($>75$\%); purple is lowest ($<25$\%); yellow is intermediate). TPF-I/Darwin targets are shown as black open circles. The loci of minimum masses detectable by SIM in 125 2-Dimensional visits are also shown. Table~\ref{startable} gives a subset of the most favorable targets.}
\label{synergyfig}
\end{figure*}

The science teams for SIM, TPF-C and TPF-I/Darwin have developed preferred lists of 100-250 stars optimized for their particular instrumental capabilities as described in (Traub \etal 2006): 

\begin{itemize}

\item While closer, lower mass stars maximize SIM's astrometric signal (cf. Eqn 1), more luminous stars have a larger habitable zone (scaling as $L_*^{0.5}$) that offsets their higher mass and typically greater distance and eases the astrometric search for habitable planets in such systems. Assuming a 1 $\mu$as sensitivity limit, the minimum mass planet in the habitable zone detectable by SIM is given by $M_{min}(SIM) = 0.33 M_\oplus d L^{-0.26}$ for a planet at $d$ pc of orbiting a star of luminosity L (L$_\oplus$) (Traub \etal  2006).

\item With its limited angular resolution, TPF-C favors the closest stars with the larger habitable zones. Assuming a limiting contrast ratio of $10^{-10}$ or 25 mag, the minimum mass planet in the habitable zone detectable by TPF-C is given by $M_{min}(TPF-C) = 0.81 M_\oplus L^{1.5}$ (Traub \etal  2006).

\item With its nearly unlimited angular resolution but more limited sensitivity, TPF-I/Darwin is best suited to the study of later spectral types with smaller habitable zones. 

\end{itemize}

Table~\ref{startable} lists the top 25 stars suitable for joint observation by TPF-I/Darwin, TPF-C, and SIM based on information gathered from the science groups of each mission. These stars have no known companions within 10$^{\prime\prime}$ and an inner edge to their habitable zones (roughly the orbit of Venus) larger than 62 mas. Stars with ecliptic latitudes in excess of 45$^o$ are excluded due to the  need to shade the TPF-I/Darwin telescopes from the sun. Also included in the table is the angular extent of the inner edge of the habitable zone (the orbit of Venus scaled by the square root of the stellar luminosity) and any indication of an infrared excess from exo-zodiacal dust as determined by IRAS or Spitzer. This information is also portrayed graphically in Figure~\ref{synergyfig} which shows stars observable in common between SIM, TPF-C, and TPF-I/Darwin.

\begin{table*}
\tabletypesize{\scriptsize} 
{\small
\caption{Likely Targets for TPF-C, TPF-I/Darwin and SIM \label{startable}}
\begin{tabular}{rrrccccr} \\ 
& & & Spec & Dist. & Inner HZ & SIM Mass & Zodiacal \\
Hip & HD & Name & Type & (pc) & (mas) & Limit (M$_\oplus^1$) & Emission \\
\tableline 
8102 & 10700 & $\tau$ Ceti & G8V & 3.65 & 120 & 0.97 & Yes, IRAS \\
19849 & 26965 & DY Eri & K0/1V & 5.04 & 75 & 1.43 & No 24/70, Nearby$^2$ \\
99240 & 190248 & $\delta$ Pav& G6/8IV & 6.11 & 116 & 1.29 & No 24/70, FGK$^2$ \\
64924 & 115617 & 61 Vir & G5V & 8.53 & 68 & 1.99 & Strong 70, FGK \\
64394 & 114710 & $\beta$ Com & G0V & 9.15 & 88 & 1.81 & No 24/70, FGK \\
15457 & 20630 & $\kappa$ Cet & G5V & 9.16 & 65 & 2.1 & N/A, FGK \\
108870 & 209100 & $\epsilon$ Ind & K4/5V & 3.63 & 60 & 1.35 & N/A, FGK \\
57443 & 102365 & GL 442A & G3/5V & 9.24 & 64 & 2.13 & No 24/70, SIMTPF$^2$ \\
14632 & 19373 & $\iota$ Per & G0V & 10.53 & 96 & 1.88 & N/A, FGK \\
12777 & 16895 & 13 Per & F7V & 11.23 & 96 & 1.95 & No 24/70, SIMTPF \\
53721 & 95128 & 47 UMa & G0V & 14.08 & 61 & 2.72 & No 24/70, FGK \\
47592 & 84117 & GL 364 & G0V & 14.88 & 65 & 2.72 & No 24/70, FGK \\
56997 & 101501 & 61 Uma & G8V & 9.54 & 52 & 2.4 & No 24/70, FGK \\
22449 & 30652 & $\pi^3$ Ori & F6V & 8.03 & 147 & 1.34 & N/A, FGK \\
78072 & 142860 & $\gamma$ Ser & F6V & 11.12 & 108 & 1.83 & No 24/70, FGK \\
25278 & 35296 & GL 202 & F8V & 14.66 & 63 & 2.74 & No 24/70, FGK \\
16852 & 22484 & GL 147 & F8V & 13.72 & 87 & 2.27 & N/A, FGK \\
80337 & 147513 & GL 620.1A & G3/5V & 12.87 & 52 & 2.8 & No 24/70, SIMTPF \\
57757 & 102870 & $\beta$ Vir & F9V & 10.9 & 120 & 1.73 & No 24/70, FGK \\
7513 & 9826 & $\upsilon$ And & F8V & 13.47 & 95 & 2.15 & N/A, FGK \\
3909 & 4813 & GL 37 & F7V & 15.46 & 59 & 2.91 & No 24/70, SIMTPF \\
116771 & 222368 & $\iota$ Psc & F7V & 13.79 & 95 & 2.19 & N/A, FGK \\
71284 & 128167 & $\sigma$ Boo& F3V & 15.47 & 83 & 2.49 & N/A, Dirty Dozen$^2$ \\
86796 & 160691 & $\mu$ Ara & G3IV/V & 15.28 & 57 & 2.93 & No 24/70, FGK \\
40843 & 69897 & $\chi$ Cnc & F6V & 18.13 & 60 & 3.13 & N/A, FGK \\
\end{tabular}
\tablenotetext{1}{Assumes 125 2-Dimensional Visits}
\tablenotetext{2}{Zodiacal emission at 24 or 70 $\mu$m in Nearby Stars MIPS Survey, Gautier \etal  (2006); FGK survey (Beichman \etal 2005a, 2006a; Bryden \etal 2006); or the SIMTPF Comparative Planetology survey (Beichman 2006b); MIPS/IRS survey of the ``Dirty Dozen'', bright, potentially resolvable disks, Stapelfeldt \etal (2006). 'N/A' denotes data not yet available in a given survey.}
}
\end{table*}

\bigskip
\noindent
\textbf{7.2 Zodiacal Dust and Planet Detection }
\bigskip

Planetary systems include many constituents: gas-giant planets, ice-giant planets and rocky planets as well as comet and asteroid belts. Understanding the interrelated evolution of all these constituents is critical to understanding the astronomical context of habitable planets and life. For example, the presence of a large amount of zodiacal emission from the debris associated with either a Kuiper Belt of comets or a rocky zone of asteroids may indicate conditions hostile to the habitable planets due to a potentially high rate of bombardment. At the same time, the transfer of water and other volatile (organic) species from the outer to the rocky planets in the habitable zone may be an essential step in the formation of life. Thus, from a scientific standpoint, we want to gather information on disks at all stages in the evolution of planetary systems, including debris disks surrounding nearby stars. TPF-C, TPF-I/Darwin, and JWST will work in conjunction to make images and spectra of scattered light and thermally emitted radiation from a large number of targets, spanning distant stars (25-150 pc) with bright disks where planets may still be forming to nearby systems with zodiacal clouds no brighter than our own. 

From an engineering standpoint, zodiacal dust is a critical factor for direct detection of planets due to increased photon shot noise and potential confusion with zodiacal structures. Sensitivity calculations for various TPF-I/Darwin and TPF-C designs suggest that the integration time needed to reach a certain level increases by a factor of 2-3 for zodiacal levels roughly 10 times the solar system's. Spitzer is carrying out a number of programs to assess the level of exo-zodiacal emission. Initial results suggest that only 15\% of solar type stars have more than 50 times the solar system's level of zodiacal emission at 30 $\mu$m, corresponding to material just outside the habitable zone, beyond about 5 AU (Beichman \etal 2006a; Bryden \etal 2006). This result is encouraging, but must be expanded to more TPF target stars using Spitzer and Herschel, and to lower levels of zodiacal emission using interferometric nulling at 10 $\mu$m on the Keck and LBT interferometers. The combination of these results will yield the ``luminosity function'' of disks for statistical purposes and allow us to screen potential targets. As an example of the sort of problem that can arise is the remarkable star HD 69830, a 2-4 Gyr old K0 star at 14 pc that might be a TPF target except for a zodiacal dust level in the habitable zone that is 1,400 times higher than seen in the solar system (Beichman \etal 2005b). While SIM may be able to identify planets around this star astrometrically, no direct searches of the habitable zone will be possible.

\section{DISCUSSION AND CONCLUSIONS}

The exploration of extrasolar planetary systems is a rich and diverse field. It calls for measurements with many kinds of instruments, as well as theoretical studies and numerical modeling. To discover and characterize extrasolar planets that are habitable and to be sure beyond a reasonable doubt that we can detect life, we need to measure the statistical distribution of planet diameters, the masses of nearby planets, and the spectra at visible and infrared wavelengths. The missions that can carry out these measurements are COROT/Kepler, SIM, TPF-C, and TPF-I/Darwin. JWST will provide followup observations for the transit missions as well as make important measurements of protostellar and debris disks. Each of these missions is a vital element of the program. Not only does each mission by itself provide its own compelling science, but together these missions form a coherent approach that will advance our understanding better than any single one by itself.

The exciting scientific promise described in preceding sections will not happen cheaply or overnight. The Great Observatories program for astrophysics spanned more than a decade between the first and last launches --- HST in 1990 and Spitzer in 2003 --- and was still longer in gestation. While the transit missions COROT and Kepler are being readied for flight, the ``Great Observatories'' for planet finding will take a generation of scientific and political advocacy: SIM PlanetQuest has completed its technology development, has been endorsed repeatedly by the US science community, and awaits final NASA approval; TPF-C and TPF-I/Darwin are well along in their programs of technology development and will need strong endorsement by US and European scientific communities in coming years. Interest is TPF goals is growing in Japan as well (Tamura \& Abe 2006). JWST is moving into its construction phase and will provide many observations useful to the planet finding endeavour. 

While the study of extrasolar planets is a new field of research with a relatively small number of (young) practitioners, our field is growing rapidly. We will fare well in any assessment of the importance of our field to progress in astronomy and of the great interest our program holds for the general public. We will also fare well in any assessment of the value of these planet finding facilities for general astrophysical investigations: SIM has already allocated a significant amount of observing time to a broad suite of general astrophysics; TPF-C will provide wide-field optical imaging with unprecedented sensitivity and angular resolution to complement JWST's mid- and near-IR capabilities; TPF-I/Darwin will break new ground with milliarcsecond mid-IR imaging and micro-Jy sensitivity. 

Near-term political considerations must not discourage us as we plan this new era of planetary exploration. From the funding agencies, we must demand the budgets needed to nurture young scientists and senior researchers, to prepare the difficult technologies of nulling, large space optics, and formation flying, and eventually to build these ``Great Observatories.'' With our colleagues, we must argue forcefully for a balanced program based on scientific priorities free from parochial considerations of individual facilities or institutions. 20$^{th}$ Century cosmologists expanded our conception of the Universe with the discovery of galaxies, the expanding universe, and dark matter. 21$^{st}$ Century planet finders will expand our conception of humanity's place in the Universe with the discoveries of other habitable worlds and possibly of life itself.

\section{ACKNOWLEDGMENTS}

We would like to thank Peter Lawson for creating Figure \ref{projfig} and Oliver Lay for his work on putting together Figure~\ref{synergyfig} and Table~\ref{startable}. The science teams of SIM, TPF-C, and TPF-I, and in particular Jim Kasting, Mike Shao, and Ken Johnston, helped develop the synergy arguments presented in $\S$6. Steve Unwin, who has been unstinting in his efforts on behalf of SIM and TPF over many years, helped to organize this special session on planet finding from space.  We are grateful to Bo Reipurth for his heroic efforts in putting together this wonderful meeting. Some of the research described in this publication was carried out at the Jet Propulsion Laboratory, California Institute of Technology, under a contract with the National Aeronautics and Space Administration.

\centerline\textbf{ REFERENCES}
\bigskip
\parskip=0pt
{\small
\baselineskip=11pt

\refs Angel, J. R. P. and Woolf, N. 1997, \apj, 475, 373.

\refs Baglin, A. 2003, Adv. Space Res, 31, 345.

\refs {{Barnes}, J.~W. and {Fortney}, J.~J.}, 2003, \apj, 588, 545.

\refs Beichman, C.A., 1998, in {\it Exozodiacal Dust Workshop}, eds. Dana Backman and Larry Caroff (NASA Technical Report), p. 149.

\refs Beichman, C. 2000, in {\it Planetary Systems In The Universe: Observation, Formation And Evolution}, eds. Alan Penny, Pawel Artymowicz, Anne-Marie LaGrange and Sara Russell, in press. 

\refs Beichman, C. et al. 2005a, \apj, 622, 1160. 

\refs Beichman, C. et al. 2005b, \apj, 626, 1061. 

\refs Beichman, C. et al. 2006a, \apj, in press.

\refs Beichman, C. et al. 2006b, in preparation.

\refs {Bord\'e}, P., {Rouan}, D. and {L{\'e}ger}, A., 2003, \aap, 405, 1137. 

\refs Borucki, W. et~al. 2003, SPIE, 4854, 129.

\refs Bracewell, R., 1978, {\it Nature}, 274, 780. 

\refs {Brown}, T.~M. and {Charbonneau}, D., {Gilliland}, R.~L., {Noyes}, R.~W. and {Burrows}, A. 2001, \apj, 552, 699.

\refs Brown, R. A. 2005, \apj,  624, 1010.

\refs  Bryden, G. et al 2006, \apj, in press.

\refs Charbonneau, D. et~al. 2005, \apj, 626, 523.

\refs Chelli, A. 2005, \aap, 441, 1205.

\refs Dekany R., Stapelfeldt K., Traub W. Macintosh B.,Woolf N., Colavita M., Trauger J. and Ftaclas, C. 2006, PASP, submitted.
 
\refs {Deming}, D. and {Seager}, S. and {Richardson}, L.~J. and {Harrington}, J. 2005, \nat, 434, 740.

\refs Gautier, T. N. et al. 2006, in preparation.

\refs Gilliand, R. et~al., 2000, \apj, 545, L47.

\refs {Gillon}, M. and {Courbin}, F. and {Magain}, P. and {Borguet}, B., 2005, 
\aap, 442, 731-744.

\refs Green, J. et al. 2005a, Proceedings of the SPIE, 5170, 38.

\refs Green, J. et al. 2005b, in {\it Techniques and Instrumentation for Detection of Exoplanets II}, Editor, Daniel R. Coulter, Proceedings of the SPIE, 5905, in press.

\refs Guyon, O., Pluzhnik, E. A., Galicher, R., Martinache, F., Ridgway, S. T., Woodruff, R. A. 2005, \apj, 622, 744.

\refs Herbst, T. M. and Hinz, P. M. in {\it New Frontiers in Stellar Interferometry}, Proceedings of SPIE Volume 5491. Edited by Wesley A. Traub. Bellingham, WA: The International Society for Optical Engineering, 2004., p.383

\refs Krist, J.E. et al. 2004 {\it High-Contrast Imaging With The Hubble Space Telescope:  Performance And Lessons Learned} in {\it Optical, Infrared, and Millimeter Space Telescopes}, Proc. SPIE, 5487, 1284 

\refs Kuchner, M. J. and Traub, W. A., 2002, \apj, 570, 900. 

\refs Lawson, P., Unwin, S., Beichman, C. A. 2004, {\it Terrestrial Planet Finder Precursor Science}, JPL Technical Report, \#04-014, 
http://planetquest.jpl.nasa.gov/documents/RdMp273.pdf

\refs Lindegren L. and Perryman M.A.C. 1996, {\it A\&AS}, 116, 579.

\refs Marcy G.W., Butler, R.P., Fischer, D.A., Laughlin, G., 
Vogt, S. S., Henry, G. and Pourbaix, D, 2002, \apj, 581, 1375.

\refs Marcy, G.W., Butler, R.P., Fischer, D., Vogt, S., Wright, J.T., Tinney, 
 C. G. and Jones, Hugh R. A. 2005, {\it Progress of Theoretical Physics Supplement}, 158, 1

\refs Martin, S.R., Gappinger, R.O., Loya, F.M., Mennesson, B.P., Crawford, S.L., Serabyn, E. 2003, {\it Techniques and Instrumentation for Detection of Exoplanets.}, Edited by Coulter, Daniel R., Proceedings of the SPIE, 5170, 144.

\refs Mennesson, B. P., Johnston, K. J., Serabyn, E. 2004, 
{\it New Frontiers in Stellar Interferometry}, Proc. SPIE, ed. Wesley A. Traub. Bellingham, WA: The International Society for Optical Engineering, 5491, p.136

\refs Moutou, C. et~al. 2005, \aap, 437, 355.

\refs Muterspaugh, M. W., Lane, B. F., Konacki, M., Burke, B. F., Colavita, M. M., Kulkarni, S. R., Shao, M. 2005, \aj, 130, 2866.

\refs Perryman, M. A. C., de Boer K.S., Gilmore G., H\o{}g E., Lattanzi M.G., et al.\ 2001.\ GAIA: composition, formation and evolution of the Galaxy.\ {\it A\&A}, 369, 339

\refs Quirrenbach, A.\ 2000.\ Astrometry with the VLT Interferometer.\ In
{\it From Extrasolar Planets to Cosmology: The VLT Opening Symposium, } ed.\ J
Bergeron, A Renzini, pp.\ 462-67.\ Berlin/Heidelberg: Springer-Verlag.

\refs Quirrenbach, A.\ 2002.\ {\it The Space Interferometry Mission
(SIM) and Terrestrial Planet Finder (TPF).} In {\it From Optical to Millimetric
Interferometry: Scientific and Technological Challenges.} Eds.\ Surdej, J., 
Swings, J.P., Caro, D., \& Detal, A., Universit\'e de Li\`ege, pp.\ 51-67.

\refs Quirrenbach, A, Coud{\'e} du Foresto V, Daigne G, Hofmann KH, 
Hofmann R, et al.\ 1998.\ PRIMA--A Study For a Dual-Beam Instrument for the
VLT Interferometer.\ In {\it Astronomical interferometry, } ed.\ RD Reasenberg, 
pp.\ 807-17.\ SPIE Vol.\ 3350.\ Bellingham, WA.

\refs Rivera, E., Lissauer, J., Butler, R.P., Marcy, G.W., Vogt, S. 
Fischer, D.A., Brown, T. \& Laughlin, G. 2005, \apj, in press.

\refs Rowe, J. \etal 2006, {\it Bull. Am. Astr. Soc},207, 110.07.

\refs Sahu, K., et al. 2005, HST Proposal ID\#10466, \break
http://adsabs.harvard.edu/cgi-bin/nph-bib\_query?bibcode= \break 2005hst..prop.6787S\&db\_key=AST

\refs Schneider, G. 2002 {\it Domains of Observability in the Near-Infrared With
HST NICMOS and Large Ground-based Telescopes}, http://nicmos.as.arizona.edu:8000/REPORTS/

\refs {Seager}, S. and {Hui}, L. 2002, \apj, 574, 1004.

\refs Serabyn, E. et~al. in {\it New Frontiers in Stellar Interferometry}, Proceedings of SPIE Volume 5491. Edited by Wesley A. Traub. Bellingham, WA: The International Society for Optical Engineering, 2004., 806.

\refs Shao, M. and Colavita, M.M., 1992, {\it A\&A} 262, 353.

\refs Shao, M.,  Wallace, J. K., Levine, B. M. and Liu, D. T. 2004,  
Proceedings of the SPIE, 5487, 1296. 

\refs Stapelfeldt, K. R. et al. 2006, in preparation.

\refs Tamura, M. and Abe, L. 2006, presented at IAU Colloquium \#200, Nice, France, in press.

\refs Traub, W.A., Carleton, N.P.,  and Porro, I.L. 1996, {\it Journal of Geophysical Research}, 101, 9291.
  
\refs Traub, W. A. and Vanderbei, R.J., 2005, \apj, 626, 1079.

\refs Traub, W.A, Kasting, J., Shao, M., Johnston, K. and Beichman, C. A. 2006, in preparation.

\refs  Unwin, S. C. 1995, {\it Astrometry in the Age of the Next Generation of Large Telescopes}, ASP Conference Series, Vol. 338, Edited by P. Kenneth Seidelmann and Alice K. B. Monet. San Francisco: Astronomical Society of the Pacific, 2005., p.37

\refs Vanderbei, R. J., Kasdin, N. J., Spergel, D. N. 2004, \apj, 65, 555. 

\refs Walker G. \etal 2005, \apj, 635, L77.
}



\end{document}